\documentclass[11pt]{article}
\usepackage{graphicx}
\begin{document}
\begin{large}
\title{\bf {Origin of which-way information and generalization of the Born rule}}
\end{large}
\author{Bruno Galvan \footnote{Electronic address: b.galvan@virgilio.it}\\ \small Loc. Melta 40, 38014 Trento, Italy.}
\date{\small May 2007}
\maketitle

\begin{abstract}
The possibility to recover the which-way information, for example in the two slit experiment, is based on a natural but implicit assumption about the position of a particle {\it before} a position measurement is performed on it. This assumption cannot be deduced from the standard postulates of quantum mechanics. In the present paper this assumption is made explicit and formally postulated as a new rule, the {\it quantum typicality rule}. This rule correlates the positions of the particles at two different times, thus defining their trajectories. Unexpectedly, this rule is also equivalent to the Born rule with regard to the explanation of the results of statistical experiments. For this reason it can be considered a generalization of the Born rule. The existence of the quantum typicality rule strongly suggests the possibility of a new trajectory-based formulation of quantum mechanics. According to this new formulation, a closed quantum system is represented as a {\it quantum process}, which corresponds to a canonical stochastic process in which the probability measure is replaced by the wave function and the usual frequentist interpretation of probability is replaced by the quantum typicality rule.
\end{abstract}
\section{Introduction} \label{introduction}
Let us consider the following very simple experiment:
\begin{center}
\includegraphics {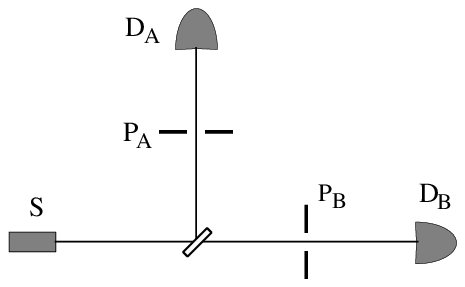} \\
Fig. 1
\end{center}
The source $S$ emits photons towards a beam splitter, and the reflected (transmitted) photons are detected by the detector $D_A$ ($D_B$). A very natural assumption is that the photons detected by $D_A$ ($D_B$) have crossed the pinhole $P_A$ ($P_B$). From this assumption, one can deduce the so-called {\it which-way information}, which is usually considered in discussions of the complementarity principle and in analysis of the delayed choice experiments \cite{bohr, wheeler}. This assumption refers to the position of a particle {\it before} a position measurement is performed on it and, in spite of evidence for its validity, it cannot be deduced from the standard postulates of quantum mechanics. Indeed, according to the standard formulation, the wave function of any photon is split into two wave packets by the beam splitter, and both wave packets exist until one of them is detected by a detector and the other collapses. If the apparatus is large enough, according to this representation (the wave function of) any photon crosses both pinholes. Why, then, do we claim on the contrary that the photons detected, for example, by the detector $D_A$ have crossed the pinhole $P_A$?

We will see that in order to obtain such a natural conclusion, it is necessary to postulate a new rule, analogous to --but different from-- the Born rule. For reasons that will be made clear in section \ref{sec:comparison}, this rule will be referred to as the {\it quantum typicality rule}. This rule correlates the position of a particle (or of a system of particles) at two different times, thus defining trajectories for the particle, although in an approximate way.

The existence of the quantum typicality rule strongly suggests the possibility of a new formulation of quantum mechanics. According to such a formulation, the particles follows definite trajectories, as in Bohmian mechanics, but the trajectories are defined by the quantum typicality rule instead of by a guidance equation as in Bohmian mechanics \cite{bohm1, bendl, allori, durr1}. Moreover, the quantum typicality rule will be found to be equivalent to the Born rule, with regard to the explanation of the results of statistical experiments. For this reason, the quantum typicality rule can be considered a generalization of the Born rule. In conclusion, the quantum typicality rule alone is sufficient to determine the properties, either dynamic or statistical, of the trajectories.

In section \ref{qtr} the quantum typicality rule will be formally expressed. In section \ref{unruh} it will be applied, as an example, to the analysis of an interferometry experiment. In section \ref{formulation} the new formulation of quantum mechanics based on the quantum typicality rule will be expressed in a formal way and discussed. 

Parts of this paper have been yet presented in a previous paper \cite{galvan}.

\section{The quantum typicality rule} \label{qtr}
Let us explicitly formulate the assumption underlying the possibility to recover the which-way information. Let $\Psi(t)=U(t)\Psi_0$ be the wave function of a particle, where $U(t)$ is the unitary time evolution operator, and $\Psi_0$ is the normalized wave function of the particle at the time $t=0$. Let us suppose that at a time $t_1$ the wave function can be expressed as the sum of two non-overlapping wave packets $\phi$ and $\phi_\perp=\Psi(t_1)-\phi$, and that at a time $t_2 > t_1$ the two wave packets are still non-overlapping, i.e. $U(t_2-t_1)\phi$ and $U(t_2-t_1)\phi_\perp$ are non-overlapping. With respect to the experiment described in the previous section, the wave packet $\phi$ is, for example, the transmitted wave packet at the time $t_1$ when it crosses the pinhole $P_A$, and $U(t_2-t_1)\phi$ is the same wave packet at the time $t_2$ when the photon is revealed by the detector $D_A$. The assumption is as follows: if the particle is found inside the support of $U(t_2-t_1)\phi$ at the time $t_2$, then it was also inside the support of $\phi$ at the time $t_1$, {\it even if no measurement has been performed at the time $t_1$}.

Although this assumption is very reasonable, it cannot be derived from the standard postulates of quantum mechanics, and therefore it must be postulated as a new rule. This rule appears to be empirically irrelevant, since it makes a prediction about something that, by definition, is not measurable, i.e. the position of the particle at the time $t_1$. This is true when the rule is applied to the usual microscopic systems of quantum experiments, for which the macroscopic environment is assumed as given (at least in the standard interpretations, such as the Copenhagen interpretation). However, when applied to macroscopic systems, such as the universe, the rule can explain the emergence of a quasi-classical world. This property will be discussed in detail in section \ref{formulation}. In this section, we limit ourselves to expressing the rule in a more precise and general way.

\vspace{3mm}
The condition that $\phi$ and $U(t_2-t_1)\phi$  do not overlap $\phi_\perp$ and $U(t_2-t_1)\phi_\perp$, respectively, implies that two subsets $\Delta_1$ and $\Delta_2$ of the configuration space of the particle exist, such that
\begin{equation} \label{cc1}
\phi \approx E(\Delta_1)\Psi(t_1) \; \; \hbox{and} \; \; U(t_2-t_1)\phi \approx E(\Delta_2)\Psi(t_2),
\end{equation}
where $E(\cdot)$ is the projection-valued measure on the configuration space of the particle. The sets $\Delta_1$ and $\Delta_2$ can be considered as the supports of $\phi$ and $U(t_2-t_1)\phi$ respectively. The conditions (\ref{cc1}) can be combined to give the condition
\begin{equation} \label{cc2}
U(t_2-t_1)E(\Delta_1)\Psi(t_1) \approx E(\Delta_2)\Psi(t_2).
\end{equation}
This reasoning can also be reversed: given two subsets $\Delta_1$ and $\Delta_2$ satisfying condition (\ref{cc2}), the wave packet $\phi:= E(\Delta_1)\Psi(t_1)$ satisfies the conditions of (\ref{cc1}).

Let us now take a further step. Given a time $t_i$ and a subset $\Delta_i$, for $i=1, 2$, let us introduce the notations $S_i:=(t_i, \Delta_i)$ and $\hat{S}_i:=U^\dagger(t_i)E(\Delta_i)U(t_i)$. We will state that the particle is in $S_i$ if it is in $\Delta_i$ at the time $t_i$. With this notation, condition (\ref{cc2}) becomes 
\begin{equation} \label{cc3}
||\hat{S}_1 \Psi_0 - \hat{S}_2\Psi_0||^2 \approx 0,
\end{equation}
where the norm has been squared for consistency with the Born rule, as explained in section \ref{sec:born}. Since the equality is approximated, the condition (\ref{cc3}) must be normalized. Let us consider the two following possible normalizations:
\begin{equation} \label{norm}
M_\Psi(S_1, S_2):=\frac{||\hat{S}_1 \Psi_0 - \hat{S}_2\Psi_0||^2}{\max\{||\hat{S}_1\Psi_0||^2, ||\hat{S}_2\Psi_0||^2\}}; \; \; m_\Psi(S_1, S_2):=\frac{||\hat{S}_1 \Psi_0 - \hat{S}_2\Psi_0||^2}{\min\{||\hat{S}_1\Psi_0||^2, ||\hat{S}_2\Psi_0||^2\}}.
\end{equation}
We have the inequalities
\begin{equation} \label{inem}
\sqrt{M_\Psi(S_1, S_2)} \leq \sqrt{m_\Psi(S_1, S_2)} \leq \frac{ \sqrt{M_\Psi(S_1, S_2)}}{1 -  \sqrt{M_\Psi(S_1, S_2)}},
\end{equation}
from which it follows, for example, that if $M_\Psi(S_1,S_2) \leq 0.08$, then 
$$
M_\Psi(S_1, S_2) \leq m_\Psi(S_1,S_2) \leq 2 M_\Psi(S_1, S_2).
$$
This means that $M_\Psi(S_1, S_2) \ll 1 \Leftrightarrow m_\Psi(S_1, S_2) \ll 1$. By normalizing condition (\ref{cc3}), we obtain the conditions $M_\Psi(S_1, S_2) \ll 1 $ or $m_\Psi(S_1, S_2) \ll 1$, depending on the normalization chosen, so that the two normalizations are equivalent for our purpose. For reasons that will be explained in section \ref{sec:born}, the normalization $M_\Psi$ will be chosen.

In conclusion, we can postulate the 
\begin{trivlist}
\item[\hspace\labelsep{\bf Quantum Typicality Rule:}] let $\Psi(t)$ be the wave function of a quantum particle (or of a system of particles). If the particle is in $S_2$, and 
\begin{equation}
M_\Psi(S_1, S_2):=\frac{||\hat{S}_1 \Psi_0 - \hat{S}_2\Psi_0||^2}{\max\{||\hat{S}_1\Psi_0||^2, ||\hat{S}_2\Psi_0||^2\}} \ll 1,
\end{equation}
than (almost certainly) it was or will be in $S_1$, and vice-versa.
\end{trivlist}
The reason for the name {\it quantum typicality rule} will be made clear in section \ref{sec:comparison}. Note that the condition $t_1 < t_2$ has not been included in the formal definition of the rule, which is time-symmetric. The use of the expression ``the particle is'' in place of ``the particle is found'' suggests the view that the particles have a definite position, independently of any measurement. This is a natural consequence of the quantum typicality rule and will be formalized in section \ref{formulation}. The meaning of the words ``almost certainly'' will be made clearer in section \ref{sec:comparison}.

\section{Analysis of the Unruh experiment} \label{unruh}

Let us now apply the quantum typicality rule to the analysis of an experiment more complex than that proposed in the introduction. This experiment has been recently proposed by Unruh in order to support the complementarity principle against possible criticism \cite{afshar, unruh}. However, here we are interested in an exemplification of the quantum typicality rule and not in a discussion of the complementarity principle.

Figure 2 shows a multiple pass Mach-Zender interferometer:
\begin{center}
\includegraphics {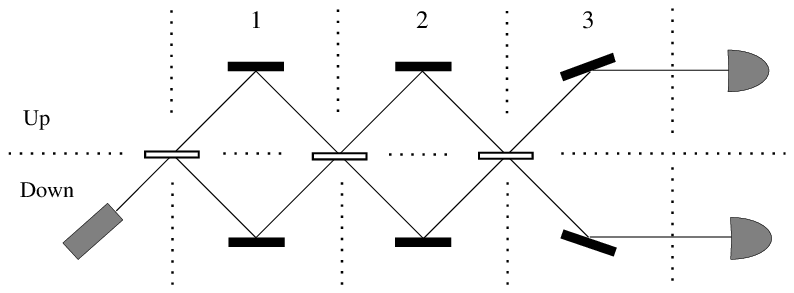} \\
Fig. 2
\end{center}
A source (on the left) emits photons towards a system of full- and half-silvered mirrors, with two detectors at the end (on the right). The half-silvered mirrors are arranged along the middle line. For simplicity, let us assume that the configuration space of the photon is the bi-dimensional plane of the figure. The line of the half-silvered mirrors divides the plane into the upper half-plane $\Delta_U$ and the lower half-plane $\Delta_D$. Moreover, let $\Delta_i$, for $i=1, 2,3$, denote the vertical sections of the plane as shown in the figure. At the time $t_0=0$ a photon with wave function $\Psi_0$ is emitted by the source. Let $t_i$, $i=1, 2,3$, be the times at which the support of the wave function of the photon is fully contained in $\Delta_i$ (we assume that the apparatus is large enough to allow this possibility). Finally, let $U_i$ and $D_i$ denote the pairs $(t_i, \Delta_U \cap \Delta_i)$ and $(t_i, \Delta_D \cap \Delta_i)$, respectively. Thus we have, for example, that $U(t_3)\hat{D}_3\hat{U}_2\hat{U}_1\Psi_0$ is the wave packet of the photon which has been transmitted by the first half-silvered mirror, reflected by the second one and transmitted by the third one. The projections $\hat{U}_i$ and $\hat{D}_i$ satisfy the following completeness relations:
\begin{equation} \label{u1}
(\hat{U}_1 + \hat{D}_1)\Psi_0 \approx \Psi_0, \; \; (\hat{U}_2 + \hat{D}_2)\hat{X}_1\Psi_0 \approx \hat{X}_1\Psi_0 \; \; \hbox{and} \; \; (\hat{U}_3 + \hat{D}_3)\hat{Y}_2 \hat{X}_1\Psi_0 \approx \hat{Y_2}\hat{X}_1\Psi_0,
\end{equation}
where $X, Y=U, D$. At every reflection a wave packet is subjected to a phase-shift of $\pi/2$; since all the paths between two mirrors have exactly the same length, we have the following equalities:
\begin{eqnarray} \label{u2}
\hat{U}_3\hat{U}_2\hat{U}_1\Psi_0 =:  + \Psi_U;  & & 
\hat{D}_3\hat{U}_2\hat{U}_1\Psi_0 =:  + \Psi_D; \nonumber \\
\hat{U}_3\hat{U}_2\hat{D}_1\Psi_0 \approx + \Psi_U;  & & 
\hat{D}_3\hat{U}_2\hat{D}_1\Psi_0 \approx  + \Psi_D;  \\
\hat{U}_3\hat{D}_2\hat{U}_1\Psi_0 \approx - \Psi_U; & & 
\hat{D}_3\hat{D}_2\hat{U}_1\Psi_0 \approx  + \Psi_D; \nonumber \\
\hat{U}_3\hat{D}_2\hat{D}_1\Psi_0 \approx  + \Psi_U; & & 
\hat{D}_3\hat{D}_2\hat{D}_1\Psi_0 \approx - \Psi_D. \nonumber
\end{eqnarray}
From (\ref{u1}) and (\ref{u2}) we obtain
\begin{equation} \label{u3}
\hat{U}_1\Psi_0 \approx \hat{D}_3\Psi_0; \; \; \hat{D}_1\Psi_0 \approx \hat{U}_3\Psi_0; \; \; \hat{U}_2\Psi_0 \approx \Psi_0.
\end{equation}
By applying the quantum typicality rule to the first two equalities, we obtain the result that, if the photon is in $U_3$ ($D_3$), then it was also in $D_1$ ($U_1$). Moreover, from the third equality, we have the result that the photon is with certainty in $U_2$. Thus, the possible trajectories of the photon are those shown in the figure:
\begin{center}
\includegraphics {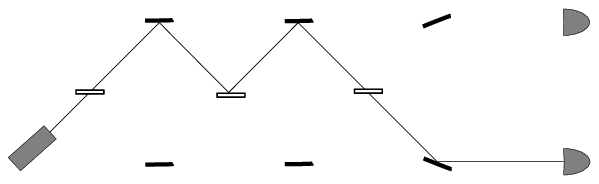} 
\includegraphics {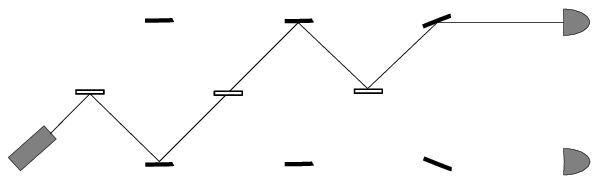} \\
Fig. 3
\end{center}
We see that the information about which of the two sectors U-1 or D-1 the photon crosses at the time $t_1$ is not destroyed at the time $t_3$, and it can be recovered by the experimenter. At the same time, we have destructive interference in the sector D-2, and no photon crosses this sector. A consequence of the equalities in (\ref{u3}) is that, if we insert an obstacle, for example in the arm U-1, the counting rate of the upper detector does not change, while the counting rate of the lower detector becomes zero. This is the usual way for deducing the which-way information.

Let us now examine the situation in which, in order to verify the presence of destructive interference, a detector in inserted in the arm D-2:
\begin{center}
\includegraphics {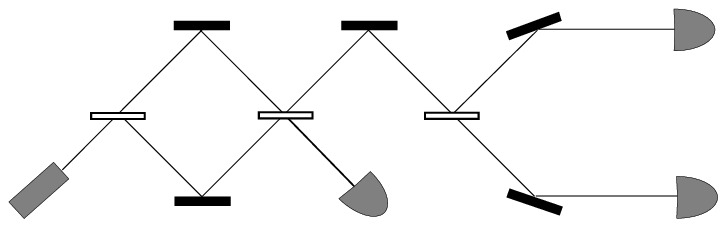} \\
Fig. 4
\end{center}
With respect to the equalities (\ref{u2}), the situation is now the following: the terms of the type $\hat{Y}_3\hat{U}_2\hat{X}_1\Psi_0$ remain unchanged, while the terms of the type $\hat{Y}_3\hat{D}_2\hat{X}_1\Psi_0$ are zero. As a consequence, the equality $\hat{U}_2\Psi_0 \approx \Psi_0$ still holds, so that no photon is detected by the counter in the arm D-2. On the contrary, there are no longer equalities of the type $\hat{X}_1\Psi_0 \approx \hat{Y}_3\Psi_0$, so that the information from the arm crossed by the photon at the time $t_1$ is destroyed at the time $t_3$. The figure shows the four admissible trajectories of the photon together (any possible path between the source and a detector along the lines in the figure is an admissible trajectory):
\begin{center}
\includegraphics {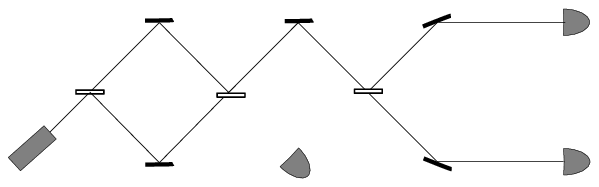} \\
Fig. 5
\end{center}

In conclusion, when the interference is measured, the which-way information is destroyed. This conclusion corresponds to that of Unruh.

\section{A new formulation of quantum mechanics} \label{formulation}
The quantum typicality rule strongly suggests a possible new formulation of quantum mechanics, already proposed in \cite{galvan}, and which is again developed here.

Let $M=R^{3N}$ be the configuration space of an N-particle system, ${\cal B}$ the $\sigma$-algebra of the Borel subsets of $M$, and $T$ a suitable time interval, for example $[0, +\infty)$. Let $M^T$ denote the set of all the trajectories $\Lambda:T \rightarrow M$. Given $t \in T$ and $\Delta \in {\cal B}$, we redefine the symbol $S=(t, \Delta)$ as the subset $\{\lambda \in M^T: \lambda(t) \in \Delta \}$. The sets of this type will be referred to as {\it single-time cylinder sets} (s-sets for short). 

As to the quantum formalism, let $\Psi$ synthetically denote the quantum structure $({\cal H}, E(\cdot), H, \Psi_0)$, where: ${\cal H}$ is a Hilbert space, $E(\cdot)$ is a projection-valued measure on $\cal B$, $H$ is the Hamiltonian, which defines the time evolution operator $U(t):=\exp[- \frac{i}{\hbar} H t]$, and $\Psi_0 \in {\cal H}$ is the wave function at the time $t=0$. As before, given the s-set $S=(t, \Delta)$, the symbol $\hat{S}$ denotes the projection $U^\dagger(t)E(\Delta)U(t)$.

The new formulation of quantum mechanics we propose is based on the following three postulates:
\begin{itemize}
\item[1.] The evolution of a closed system of $N$ non relativistic quantum particles, possibly the universe, is described by a suitable trajectory $\lambda \in M^T$.
\item[2.] A quantum structure $\Psi$ is associated with the system.
\item[3.] The dynamic and statistical properties of the trajectory $\lambda$ are determined by the quantum structure, by means of the {\it quantum typicality rule}: given two s-sets $S_1$ and $S_2$, if $\lambda$ belongs to $S_1$, and
\begin{equation}
M_\Psi(S_1, S_2) := \frac{||\hat{S}_1 \Psi_0 - \hat{S}_2\Psi_0||^2}{\max\{||\hat{S}_1\Psi_0||^2, ||\hat{S}_2\Psi_0||^2\}} \ll 1,
\end{equation}
then, almost certainly, $\lambda$ belongs to $S_2$, and vice versa.
\end{itemize}

Such a mathematical model for a quantum system will be referred to as a {\it quantum process}. Let us now examine some of its properties.

\subsection{Comparison with a stochastic process} \label{sec:comparison}

In order to clarify the conceptual meaning of the various elements of a quantum process, it is very useful to compare it with a canonical stochastic process. A canonical stochastic process is composed of the set $M^T$, endowed with a probability measure $\mu$ defined on the $\sigma$-algebra generated by the s-sets. According to the {\it Kolmogorow reconstruction theorem}, the probability measure is unequivocally determined by its {\it finite dimensional distributions}, i.e. the values of the measure on the finite intersections of s-sets:
\begin{equation}
\mu(S_1 \cap \ldots \cap S_n).
\end{equation}
We thus have the following natural correspondences:
\begin{eqnarray}
\hbox{Quantum process} \; \; & & \; \; \hbox{Stochastic process} \nonumber \\
M^T & < \! = \! > & M^T  \label{c1} \\
\Psi & < \! \approx \! > & \mu \label{c2} \\
||\hat{S}\Psi_0||^2 & <\!=\!> & \mu(S) \label{c3} \\
\hbox{the Born rule} & <\!=\!> & \hbox{the interpr. of} \; \; \mu(S) \label{c4} \\ 
M_\Psi(S_1, S_2) & < \! \approx \! > & M_\mu(S_1, S_2) \label{c5} \\
\hbox{the quantum typicality rule} & <\!=\!> & \hbox{the interpr. of} \; \; M_\mu(S_1, S_2) \ll 1 \label{c6} \\ 
||T(\hat{S_1} \ldots \hat{S_n})\Psi_0||^2 & <\!\neq\!> & 
\mu(S_1 \cap \ldots \cap S_n) \label{c7} 
\end{eqnarray}
Let us examine the various correspondences.

\vspace{3mm}
{\it Correspondence (\ref{c1})}. For both the kind of processes, the evolution of the represented physical system corresponds to a suitable trajectory $\lambda \in M^T$. This means that the formulation of quantum mechanics based on a quantum process is a {\it trajectory-based} formulation, analogous for example to Bohmian mechanics. In such formulations, the positions of the particles are the only observables, and neither the measurement process nor the observers enter into the theory on a fundamental level. The way in which the standard quantum measurement theory is derived in these formulations is presented for example in \cite{bendl}.

Since the set $M^T$ has no structure, it has no empirical content, i.e. no empirical prediction can be derived from it. This is also the case with a stochastic process, where all empirical predictions derive from the measure $\mu$ alone. However, analogously to a canonical stochastic process, the set $M^T$ is necessary to ensure the structural coherence of the model, and it cannot be eliminated. By eliminating it, one would obtain something like the Many Worlds Interpretation \cite{everett, dewitt}, with its well known conceptual and interpretative issues. We will return to this point later.

\vspace{3mm}
{\it Correspondence (\ref{c2})}. In a quantum process the quantum structure $\Psi$, in particular the wave function, plays the same role as that of the probability measure in a stochastic process, i.e. it determines the dynamic and statistical properties of the trajectories. The correspondence (\ref{c2}) is however approximate, because a quantum structure and a probability measure have certain structural differences.

\vspace{3mm}
{\it Correspondences (\ref{c3}) and (\ref{c4})}. From the mathematical point of view, correspondence (\ref{c3}) is exact and self-evident, because, for fixed $t$, both the expressions $||E(\cdot)\Psi(t)||^2$ and $\mu[(t, \cdot)]$ are probability measures of $M$. 

Let us now consider correspondence (\ref{c4}). The usual frequentist interpretation of $\mu(S)$ can be expressed as follows:
\begin{trivlist}
\item[\hspace\labelsep{\bf Interpretation of $\mu(S)$:}] if a large number $N$ of physical systems are represented by the same stochastic process with probability measure $\mu$, then $N \cdot \mu(S)$ is approximately the number of systems whose trajectories belong to $S$.
\end{trivlist}
In the context of a quantum process, the Born rule can be expressed as follows:
\begin{trivlist}
\item[\hspace\labelsep{\bf The Born rule:}] if a large number $N$ of physical systems are represented by the same quantum process with quantum structure $\Psi$, then $N \cdot ||\hat{S}\Psi_0||^2$ is approximately the number of systems whose trajectories belong to $S$.
\end{trivlist}
These two formulations make correspondence (\ref{c4}) self-evident. The reason why the expression $||\hat{S}\Psi_0||^2$ and the Born rule have not been included in the new formulation will be explained in section \ref{sec:born}.

\vspace{3mm}
{\it Correspondences (\ref{c5}) and (\ref{c6})}. In correspondence (\ref{c5}) the set function 
\begin{equation} \label{poa}
M_\mu(S_1, S_2) :=\frac{\mu(S_1 \triangle S_2)}{\max\{\mu(S_1), \mu(S_2)\}}
\end{equation}
has been introduced (the set function $m_\mu(S_1, S_2) :=\mu(S_1 \triangle S_2)/\min\{\mu(S_1), \mu(S_2)\}$ can also be introduced; $M_\mu$ and $m_\mu$ satisfy an inequality analogous to (\ref{inem}) but without the square roots). From the formal point of view, correspondence (\ref{c5}) is based on the correspondence 
\begin{equation} \label{acor}
||\hat{S_1}\Psi_0 - \hat{S_2}\Psi_0||^2 = ||T(\hat{S_1}\hat{\bar{S}_2})\Psi_0||^2 + ||T(\hat{\bar{S}_1}\hat{S_2})\Psi_0||^2 <\!\approx \!> \mu(S_1 \cap \bar{S}_2) + \mu(\bar{S}_1 \cap S_2) = \mu(S_1 \triangle S_2),
\end{equation}
where $T( \cdot)$ is the time-ordering operator. This correspondence  cannot be exact, because it is based on correspondences of the type (\ref{c7}), which, as we will see, are not appropriate. In fact, correspondence (\ref{c5}) is appropriate only in the typicality regime, i.e. when $M_\Psi(S_1, S_2)$ and $M_\mu(S_1, S_2)$ are $\ll 1$. This is the reason for the symbol $<\!\approx \!>$  in place of $<\!=\!>$.

With regard to correspondence (\ref{c6}), the probabilistic interpretation of the condition $M_\mu(S_1, S_2) \ll 1$ is exactly the quantum typicality rule, with $M_\Psi(S_1, S_2)$ replaced by $M_\mu(S_1, S_2)$. This remark makes correspondence (\ref{c6}) obvious and clarifies the meaning of the words ``almost certainly'' in the quantum typicality rule.

In the probabilistic case, if $M_\mu(S_1, S_2) \ll 1$, the s-sets $S_1$ and $S_2$ are said to be {\it mutually typical}, i.e. such that the overwhelming majority of the elements of $S_1$ also belong to $S_2$, and vice-versa. We can extend this definition to the condition $M_\Psi(S_1, S_2) \ll 1$. For this reason, the set functions $M_\Psi(S_1, S_2)$ and $M_\mu(S_1, S_2)$ will be referred to as the quantum and the probabilistic {\it mutual typicality measure} respectively, and this is the origin of the name ``quantum typicality rule''. The set function $M_\Psi(S_1, S_2)$ defines mutual typicality without the need of an underlying probability measure; the possible independence of the two notions, typicality and probability, has been pointed out in \cite{goldstein}. 

\vspace{3mm}
{\it Correspondence (\ref{c7})}. The left-hand member of this false correspondence is a tentative expression for ``quantum'' finite dimensional distributions. Indeed, according to the Born rule and the reduction postulate, it is the probability {\it of finding } the particles in $\Delta_i$ at the time $t_i$, for $i=1, \ldots, n$. Correspondence (\ref{c7}) cannot hold true, because the quantum expression is not additive: let $t_1 \leq t_2$ and $S_1 \cap S'_1=\emptyset$. Then $S_1 \cup S'_1$ is an s-set, with $(S_1 \cup S'_1)\, \hat{} = \hat{S}_1 +\hat{S}'_1$, but $||\hat{S}_2 (\hat{S}_1 + \hat{S}'_1)\Psi_0||^2 \neq ||\hat{S}_2 \hat{S}_1\Psi_0||^2+||\hat{S}_2 \hat{S}'_1\Psi_0||^2$. This is the well known paradoxical aspect of quantum interference, which precludes the representation of a quantum system as a stochastic process. This false correspondence has been shown here only to illustrate the difference between a quantum and a stochastic process, but the expression $||T(\hat{S_1} \ldots \hat{S_n})\Psi_0||^2 $ plays no role in a quantum process. \\

\subsection{The dynamic structure of the trajectories} \label{sec:dynamics}
The quantum typicality rule establishes a correlation between the positions of the particles at two different times, thus defining a dynamic structure for the trajectories, although in an approximate way. When an experiment is performed on a microscopic quantum system, the trajectory of the system between the preparation and the measurement times has no empirical relevance and is usually considered only a matter of interpretation, as in the analysis of the Unruh experiment. The structure of the trajectory becomes empirically relevant when the system is macroscopic, and the trajectory must explain the observed quasi-classical macroscopic evolution. The explanation of this empirical evidence by means of the laws of quantum mechanics alone is a well known problem.

Let us assume that a quantum process represents an idealized non relativistic universe. It is easy to see that, due to the quantum typicality rule, the trajectories follow the branches in which the universal wave function splits, for example in the presence of a measurement-like interaction. The splitting of the universal wave function into branches is considered, in a more or less explicit manner, in many formulations of quantum mechanics, in particular in Bohmian mechanics, the Many Worlds Interpretation, the Consistent Histories formulation of quantum mechanics \cite{consistent} and the theory of decoherence \cite{decoherence}. The definition of branches as non-overlapping parts of the universal wave function is present mainly in the works relating to Bohmian mechanics, for example \cite{bohm3, struyve, peruzzi}. 

Let us consider the following qualitative description of the branching process: let $\Psi(t)$ be the universal wave function, and let us assume that at the time $t_1$ the first measurement takes place, with $n$ possible outcomes. At this time, the universal wave function splits into $n$ permanently disjoined wave packets. The wave packets are disjoined, because they correspond to $n$ different positions of a macroscopic pointer, and they are permanently disjoined due to decoherence. After this, every wave packet subsequently undergoes analogous splitting, thus giving rise to a tree structure for the universal wave function. More specifically, a branch corresponds to the choice of a wave packet for every splitting.

Let $\phi$ be a wave packet for the splitting occurring at the time $t_1$. This means that $\phi(t):=U(t-t_1)\phi$ does not overlap $\phi_\perp(t):=\Psi(t)-\phi(t)$ for $t\geq t_1$. Let $\Delta_1$ and $\Delta_2$ be the supports of $\phi(t)$ at the times $t_1$ and $t_2$, respectively. As we have seen in section \ref{qtr}, since $\phi(t)$ does not overlap $\phi_\perp(t)$ at the times $t_1$ and $t_2$, we have $M_\Psi(S_1, S_2) \ll 1$ and, according to the quantum typicality rule, a trajectory belonging to $S_2$ almost certainly belongs to $S_1$ as well. Let us then assume that, at the time $t_2$, the wave packet $\phi(t_2)$ is further split, and let $\Delta'_2$ be the support of one of the wave packets of this splitting. It is of course true that $\Delta'_2 \subseteq \Delta_2$. As a consequence, a trajectory belonging to $S'_2=(t_2, \Delta'_2)$ almost certainly also belongs to $S_1$. By iterating this reasoning, we deduce that the trajectories of the universe remain, almost certainly, inside the supports of the branches of the universal wave function.

Of course the above reasoning is on a qualitative level. In order to rigorously prove the quasi-classical structure of the macroscopic evolution, one has to rigorously prove that the wave function actually has a forward tree structure and that the branches have a quasi-classical structure. Here we do not face this problem, and we limit ourselves to the argument that the Ehrenfest theorem and Mott's analysis of the cloud chamber \cite{mott} should be important tools to obtain more rigorous results in this sense.

\subsection{Statistical experiments} \label{sec:born}
In the standard formulation of quantum mechanics, the results of statistical experiments are explained by the Born rule. At the same time, the quantum typicality rule has been established as a rule connecting the positions of the particles at two different times, thus defining the dynamic, but not the statistical, properties of the trajectories. However, perhaps unexpectedly, the quantum typicality is equivalent to the Born rule with regard to the explanation of the results of the statistical experiments. This is the case, because such results can also be explained by a reasoning based on typicality instead of probability, and because the typicality defined by the Born rule corresponds to that defined by the quantum typicality rule. Let us explain.

According to the Born rule, for every value $t$, the expression $||E( \cdot ) \Psi(t)||^2$ is a probability measure on $M$. On this basis, one can define the probabilistic mutual typicality measure
\begin{equation} \label{f1}
M_{\Psi(t)}(\Delta_1, \Delta_2):= \frac{||E(\Delta_1 \triangle \Delta_2) \Psi(t)||^2}{\max\{||E(\Delta_1) \Psi(t)||^2, ||E(\Delta_2) \Psi(t)||^2\}}.
\end{equation}
But (\ref{f1}) is exactly the quantum mutual typicality measure of the two equal time s-sets $(t, \Delta_1)$ and $(t, \Delta_2)$. Moreover, note the following equality:
\begin{equation}
||E(\overline{\Delta})\Psi(t)||^2=M_\Psi[(t, M), (t, \Delta)].
\end{equation}
The possibility of obtaining this equality is the reason for the choice of the normalization $M_\Psi$ instead of $m_\Psi$ (see definitions (\ref{norm})).

Thus, the typicality defined by the Born rule for configurations at a fixed time corresponds to the typicality defined by the mutual typicality measure for equal-time s-sets. In this sense, the quantum typicality rule and the Born rule are equivalent, or more precisely, the quantum typicality rule can be considered as a generalization of the Born rule to non equal time s-sets.

In order to see that the results of the statistical experiments can also be explained by typicality, let us consider two possible representations of such an experiment: the first based on probability, and the second based on typicality. The second representation, although not usual, is however not new \cite{dewitt}.

In the first representation, the system under consideration is a laboratory  consisting of {\it one} microscopic system plus a measuring device. The device performs the measurement of an observable of the microscopic system with n possible outcomes. The experiment starts at the time $t_1$ and ends at the time $t_2$. Let $\Psi(t)$ denote the quantum state of the laboratory. We have, as usual
\begin{equation} \label{exp1}
\Psi(t_1)=\phi \cdot \Phi_0 =\left[\sum_{s=1}^n (\phi, \phi_s)\phi_s \right ] \cdot \Phi \rightarrow \Psi(t_2)= \sum_{s=1}^n (\phi, \phi_s)\phi_s \cdot \Phi_s,
\end{equation}
where $\phi$ and $\Phi_0$ are the initial states of the microscopic system and of the device, respectively, $\phi_s$, for $s=1, \ldots, n$, are the eigenstates of the measured observable, and $\Phi_s$ is the final state of the device when the $s$-th outcome has been recorded. All these state vectors are normalized. Let $\Delta_s$ denote the support of $\phi_s \cdot \Phi_s$. It is of course true that  $\Delta_s \cap \Delta_{s'}= \emptyset $ for $s \neq s'$, and $\Delta_2=\cup_s \Delta_s$, where $\Delta_2$ is the support of $\Psi(t_2)$. Moreover, we have:
\begin{equation}
||E(\Delta_s)\Psi(t_2)||^2 \approx |(\phi, \phi_s)|^2=:p_s.
\end{equation}
By applying the Born rule to this equality, we obtain the result by repeating the experiment several times that $p_s$ is the relative frequency of the $s$-th outcome, as expected.

Let us now consider the representation based on typicality. In this case, the system under consideration consists of the measuring device plus a {\it large number $N$} of microscopic systems prepared in the same initial state $\phi$. The expression (\ref{exp1}) becomes:
\begin{eqnarray}
& & \Psi(t_1)=\phi ^N \cdot \Phi = 
\left [ \sum_{s_1 \ldots s_N} (\phi, \phi_{s_1}) \ldots (\phi, \phi_{s_N})\phi_{s_1} \ldots \phi_{s_N} \right ] \cdot \Phi \rightarrow \\
& & \Psi(t_2) = \sum_{s_1 \ldots s_N} (\phi, \phi_{s_1}) \ldots (\phi, \phi_{s_N}) \phi_{s_1} \ldots \phi_{s_N} \cdot \Phi[s_1 \ldots s_N],
\nonumber
\end{eqnarray}
where $\Phi[s_1 \ldots s_N]$ is the state of the device which has recorded the outcomes $s_1 \ldots s_N$ of the $N$ measurements. Let $\Delta[s_1 \ldots s_N]$ denote the support of $\phi_{s_1} \ldots \phi_{s_N} \cdot \Phi[s_1 \ldots s_N]$. Here also, we have the result that $\Delta[s_1 \ldots s_N] \cap \Delta[s'_1 \ldots s'_N] = \emptyset $ for $(s_1 \ldots s_N) \neq (s'_1 \ldots s'_N)$. Let us introduce the functions:
\begin{eqnarray}
f(s; s_1, \ldots , s_N)& := & \frac{1}{N}\sum_{i=1}^N \delta_{s, s_i},  \\
\delta(s_1 \ldots s_N)& := & \sum_{s=1}^n (f(s; s_1, \ldots , s_N) - p_s)^2.
\end{eqnarray}
The first function gives, for every outcome $s$ and sequence $s_1 \ldots s_N$, the relative frequency of the outcome $s$ in the sequence $s_1 \ldots s_N$, and the second function is a measure of the ``distance'' of the sequence $s_1 \ldots s_N$ from exact randomness, i.e. from the situation in which the relative frequency of every outcome $s$ is exactly $p_s$. Let us moreover define
\begin{equation}
\Delta_N^\epsilon:= \! \! \! \! \! \! \! \! \! \bigcup_{
\tiny \begin{array}{c}

s_1 \ldots s_N \\
\delta(s_1 \ldots s_N) < \epsilon
\end{array}
} \! \! \! \! \! \! \! \! \!  \Delta[s_1 \ldots s_N].
\end{equation}
If $\epsilon$ is ``little'', the set $\Delta_N^\epsilon$ consists only of configurations of the laboratory at the time $t_2$, corresponding to sequences of outcomes which are close to exact randomness. One can prove \cite{dewitt} that 
\begin{equation}
||E(\bar{\Delta}_N^\epsilon) \Psi(t_2)||^2 < \frac{1}{\epsilon N}.
\end{equation}
Thus, by choosing $N$ large enough, we have $||E(\bar{\Delta}_N^\epsilon) \Psi(t_2)||^2 \ll 1$. According to the quantum typicality rule, the actual trajectory $\lambda$ almost certainly belongs to $(t_2, \Delta_N^\epsilon)$, i.e. the observed sequence $s_1 \ldots s_N$ is, almost certainly, close to randomness.

\vspace{3mm}
The above reasoning shows that the Born rule and the quantum typicality rule are equivalent with regard to the explanation of the results of the statistical experiments. There are, however, additional considerations which suggest that typicality should be considered a more fundamental notion than probability. 

The first consideration is that the representation of a statistical experiment based on typicality is more directly connected with the empirical evidence, compared with that based on probability. Indeed, the probability of an outcome cannot be directly measured, and what is actually measured in a statistical experiment is the typicality of a global outcome, consisting of a large number of elementary outcomes. The probability of the elementary outcomes is deduced from this global outcome. 

Another consideration is that the probabilistic method is no longer applicable to the case in which the system under consideration is the universe, because an ensemble of universes is obviously not available, and the Born rule then no longer applies.

Finally, it is the typicality defined by the quantum typicality rule, and not the probability defined by the Born rule, which allows us to explain the macroscopic quasi-classical dynamic structure of the trajectories.

The fundamental role of typicality in the explanation of the observable phenomena of nature has also been emphasized by Goldstein {\it et al} \cite{goldstein, durr1}.

For these reasons, we have preferred not to include the Born rule as a postulate of the new formulation and to retain the quantum typicality rule as the only basic rule of a quantum process. It is possible that further studies may suggest revising this choice.

\subsection{Comparison with other formulations} 

Let us compare the formulation of quantum mechanics proposed in this paper with Bohmian mechanics and with the Many World Interpretation.

Bohmian mechanics is the formulation closest to the present formulation. According to Bohmian mechanics, the particles of a quantum system follow one of the trajectories defined by the guidance equation:
\begin{equation}
\frac{d{\bf x}_k}{dt}=\frac{\hbar}{m_k}\hbox{Im}\frac{{\bf \nabla}_k \Psi}{\Psi}, \; \; k=1, \ldots, N.
\end{equation}
The set $\Lambda$ of the Bohmian trajectories  is endowed with the probability measure
\begin{equation} \label{bohm1}
\mu_B(\Gamma):=||E[z_t(\Gamma)]\Psi(t)||^2,
\end{equation}
where $\Gamma \subseteq \Lambda$ (see note\footnote{More correctly, $\Gamma$ belongs to $\sigma$-algebra generated by the s-sets of the process}), and $z_t:\Lambda \rightarrow M$ is the mapping defined by $z_t(\lambda)=\lambda(t)$. Due to the equivariance property of Bohmian mechanics, the definition (\ref{bohm1}) does not depend on the time. The guidance equation (potentially) explains the quasi-classical macroscopic evolution \cite{allori2}, and the measure $\mu_B$ explains the results of the statistical experiments \cite{bendl}.

Since $(\Lambda, \mu_B)$ is a probability space, $z_t$ is a random variable, and Bohmian mechanics naturally corresponds to the stochastic process  $\{z_t\}_{t \in T}$. The single-time distribution of the Bohmian process is  $\mu_B[(t, \Delta)]=||E(\Delta)\Psi(t)||^2$. Moreover, since Bohmian mechanics is deterministic, it is not difficult to see that, for any s-set $S_1=(t_1, \Delta_1)$ and for any time $t_2$, there exists only one set $\Delta_2$ (modulo a set of zero measure), such that $\mu(S_1 \triangle S_2)=0$ ($\Delta_2$ is the set $z_{t_2}\{z^{-1}_{t_1}[\Delta_1]\}$). Thus, the set function 
\begin{equation} \label{bohm2}
\mu_B(S_1 \triangle S_2)
\end{equation}
contains the dynamic information of the guidance equation. Moreover, since $\mu_B(S)= \\ \mu_B[(t, M) \triangle \bar{S}]$, it also contains the information for the single time distribution $||E(\Delta)\Psi(t)||^2$. In conclusion, we see that the situation here is similar to a quantum process, where  the set function (\ref{bohm2}) deriving from the guidance equation is replaced by the mutual typicality measure, directly deriving from the wave function. The main difference is that, for {\it every} $S_1$ and $t_2$, there exists a set $\Delta_2$ such that $\mu_B(S_1 \triangle S_2)=0$, but not such that $M_\Psi(S_1, S_2) \ll 1$.

\vspace{3mm}
With regard to the Many World interpretation (MWI), this differs from a quantum process in two main points: (i) in a quantum process the branches are explicitly (though vaguely) defined as permanently non-overlapping parts of the universal wave function; on the contrary, no explicit definition of the worlds (i.e. the branches) is given in the MWI, in which the so-called preferred basis problem \cite{barrett} appears to remain an open problem. (ii) In a quantum process there is a definite ontology for the particles, namely a trajectory $\lambda \in M^T$; this means that the branches are interpreted as descriptions of the influence of the wave function on the trajectories, and this makes the vagueness of their definition acceptable. On the contrary, in the MWI the worlds are the ontology, and a vague definition of these is in any case a problem. Moreover, since in a quantum process the particles follow a single trajectory (and therefore a single branch), the paradoxical situation of the MWI, in which all the worlds are equally real, is avoided.

\section{Summary}
This paper proposes a new formulation of quantum mechanics, based on a generalization of the Born rule. Such a rule, here called the {\it quantum typicality rule}, derives from an assumption about the position of a particle {\it before} a position measurement is performed on it. The so called ``which-way'' information is recovered on the basis of this assumption, even if the assumption is not usually expressed in an explicit way.

The quantum typicality rule has been expressed in a mathematical form and it has two principal functions: (i) it correlates the positions of a particle (or of a system of particles) at two different times, thus defining trajectories for the particle, although in an approximate way; (ii) it is equivalent to the Born rule with regard to the explanation of the results of statistical experiments.

In the new formulation, a closed quantum system --possibly even the universe-- is represented by a mathematical model called the {\it quantum process}, which is similar to a canonical stochastic process with the probability measure replaced by the wave function and the usual frequentist interpretation of probability replaced by the quantum typicality rule.

\section{Acknowledgments} \nonumber

The autor wants to thank N. Zangh\`{\i} for a useful discussion and encouragement.

\end{document}